\documentclass[12pt]{article}
\usepackage{tikz}
\usepackage[english]{babel}
\usepackage{graphicx}
\usepackage{amsmath}
\usepackage{amsthm}
\usepackage{amssymb}
\usepackage{lscape}
\usepackage{caption}
\usepackage[hyperindex,plainpages=false]{hyperref}
\usepackage{longtable}
\usepackage[left=1.1in,right=1.1in,bottom=1in,top=1in,a4paper]{geometry}

\theoremstyle{definition}
\theoremstyle{lemma}

\title{Infinitely many symmetries and conservation laws for quad-graph equations via the Gardner method}
\author{Alexander G. Rasin
\\
Department of Mathematics,\\ Bar-Ilan University,
Ramat Gan, 52900, Israel
 \\{rasin@math.biu.ac.il}}

\begin{document}
\maketitle
\begin{abstract}
The application of the Gardner method for generation of conservation laws to all the ABS equations is considered. It is shown that all the necessary information for the application of the Gardner method, namely B\"acklund transformations and initial conservation laws, follow from the multidimensional consistency of ABS equations. We also apply the Gardner method to an asymmetric equation which is not included in the ABS classification. An analog of the Gardner method for generation of symmetries is developed and applied to discrete KdV. It can also be applied to all the other ABS equations.
\end{abstract}
\section{Introduction}
The first integrable example of a partial difference equation (P$\Delta$E) goes back to \cite{Hi1}, who wrote
down an equation on a quad-graph related by a simple change of variables to the discrete KdV equation.
However, the subject has only been intensively developed in the last few years. A particular milestone
was the so-called ABS classification of integrable scalar quad-graph equations based on the
principle of "consistency on the cube" \cite{ABS}. All the equations in this classification have the discrete
analogue of a matrix Lax pair and also a "natural" auto-B\"acklund transformation \cite{At0}. Other integrability
properties (discrete analogues of scalar Lax pairs, bihamiltonian structures and infinite numbers
of conservation laws) have been studied less.

The topic of conservation laws for the P$\Delta$E has become very important recently. This started in the work of Orphanidis \cite{Or0} in which he presents conservation laws for the fully discrete sine-Gordon equation. The next step in the research of conservation laws was the introduction of a systematic computation method. The first systematic method developed was the direct method \cite{Hy0,RH0,RH1}. This method allows use of computer algebra, which makes it more constructive. But it has some disadvantages: it requires massive computations and cannot produce infinite numbers of conservation laws. The next method for computing conservation laws was proposed in \cite{RH3}. It produces infinite numbers of conservation laws by acting with a mastersymmmetry on basic conservation laws. 

A new approach to the computation of conservation laws for P$\Delta$E has appeared very recently \cite{RS1}. It is called the Gardner method and is an analog of the Gardner method for partial differential equations. The method uses a B\"acklund transformation (BT) to generate conservation laws. It seems that Gardner method is the most efficient among the methods described above.

Symmetries for P$\Delta$E were also researched recently. They first appeared as similarity constraints
for integrable lattices \cite{NP0}. Tongas \textit{et al.} pointed out
that the similarity constraints for quad-graph equations obtained
previously are equivalent to characteristics of symmetries
\cite{TTP}. Symmetries of several quad-graph equations have been found in
\cite{LP0,LPS,PTV,TN0}. 
Hydon developed a direct method for finding symmetries based on solving functional equations by creating an
associated system of differential equations that can be solved \cite{Hy1,Hy0}. In \cite{RH2} this method was applied to P$\Delta$E. There the authors compute five-point symmetries for all the ABS equations (equations from \cite{ABS}). Symmetries for P$\Delta$E were also discussed in other articles \cite{LY0,TTX}.

Our intention is to show that the Gardner method for generation of conservation laws for P$\Delta$E can be applied systematically. We apply it to all the ABS equations and to an asymmetric quad-graph equation. 
We also show that the Gardner method can be applied for symmetries. Namely, we use it to generate an infinite number of symmetries from known one.

\section{The Gardner method}
Before presenting the Gardner method for generating conservation laws for P$\Delta$E,
we review the method for the continuum Korteweg–-de Vries equation (KdV). 
The Gardner method for KdV starts with the BT. The
BT states that if $u$ solves KdV then so does $u+v_x$ where $v$ is a solution of the system 
\begin{eqnarray*}
v_x &=& \theta - 2u - v^2  \\
v_t &=& -\frac12 u_{xx} + \left(\theta+u\right)\left(\theta - 2u - v^2\right) + u_x v
\end{eqnarray*}
Here $\theta$ is a parameter. 
It is straightforward to check that $v$ is actually the $G$ component
of a conservation law. Specifically, we have   
$$ 
\partial_t v + \partial_x \left( \frac12 u_x - (u+\theta)v  \right)   
=0\ . $$
But $v$ depends on the parameter $\theta$. Expanding in a suitable series in $\theta$ yields an infinite
number of conservation laws. For KdV the appropriate series is
$$ 
v = \theta^{1/2}
- \frac{u}{\theta^{1/2}}
+ \frac{u_x}{2 \theta}
- \frac{u_{xx}+2u^2}{4\theta^{3/2}}
+ \frac{u_{xxx}+8uu_x}{8\theta^{2}}
- \frac{u_{xxxx}+8u^3+10u_x^2+12uu_{xx}}{16\theta^{5/2}}
+ O\left(\theta^{-3}\right)\ .
$$

In \cite{RS1} the implementation of the Gardner method to the dKdV equation was given. 

Here we apply the Gardner method to the following asymmetric quad-graph equation \cite{NRGO}
\begin{equation}
\alpha u_{0,0}u_{1,0}-\beta(u_{0,0}u_{0,1}+u_{1,0}u_{1,1})=0.\label{as0}
\end{equation}
Here $k,l\in \mathbb{Z}^2$ are independent variables
and $u_{0,0}=u(k,l)$ is a dependent variable that is defined on the
domain $\mathbb{Z}^2$. We denote the values of this variable on
other points by $u_{i,j}=u(k+i,l+j)=S_k^iS_l^ju_{0,0}$, where $S_k,~S_l$ are the unit forward shift operators in $k$ and $l$ respectively. 
Equation (\ref{as0}) is not included in the ABS list, since it is asymmetric.
A BT for (\ref{as0}) is
\begin{align}
\alpha (u_{0,0}u_{1,0}+\tilde{u}_{0,0}\tilde{u}_{1,0})&-\theta(u_{0,0}\tilde{u}_{0,0}+u_{1,0}\tilde{u}_{1,0})=0,\label{in1}\\
\theta u_{0,0}\tilde{u}_{0,0}&-\beta(u_{0,0}u_{0,1}+\tilde{u}_{0,0}\tilde{u}_{0,1})=0.\label{in2}
\end{align}
Here $\theta$ is a parameter. 
This BT follows from the multidimensional consistency of equations (\ref{as0},\ref{in1},\ref{in2}) embedded in three dimensions. Let us introduce the third variable $m$ so that \begin{equation}
u_{k,l,m}=u_{0,0,0}=u_{0,0},~~~u_{0,0,1}=\tilde{u}_{0,0},~~~u_{1,0,1}=\tilde{u}_{1,0},~etc.\label{no1}
\end{equation}
Then, equations (\ref{as0},\ref{in1},\ref{in2}) can be written as
\begin{eqnarray}
P_1&:&\alpha u_{0,0,0}u_{1,0,0}-\beta(u_{0,0,0}u_{0,1,0}+u_{1,0,0}u_{1,1,0})=0,\nonumber\\
P_2&:&\alpha (u_{0,0,0}u_{1,0,0}+u_{0,0,1}u_{1,0,1})-\theta(u_{0,0,0}u_{0,0,1}+u_{1,0,0}u_{1,0,1})=0,\label{sys10}\\
P_3&:&\theta u_{0,0,0}u_{0,0,1}-\beta(u_{0,0,0}u_{0,1,0}+u_{0,0,1}u_{0,1,1})=0.\nonumber
\end{eqnarray}
Equations $P_2$ and $P_3$ form a BT for $P_1$. In the sequel we use the notation (\ref{no1}), writing $u_{0,0,1}$ and $u_{0,0,-1}$ for the BT and inverse BT of $u_{0,0,0}$, where possible, however, we revert to 2 dimensional notation.

In general we cannot solve the equations of the BT to write $u_{0,0,1}$ in
terms of $u_{0,0,0}$. However there is a special case $\theta=\alpha$ for which $u_{0,0,1}=u_{1,0,0}$ or $u_{-1,0,0}$. Consider $\theta=\alpha+\epsilon$ where $\epsilon$ is small, and look for a solution of BT in the form
\begin{equation}
u_{0,0,1}=u_{1,0}+\sum_{i=1}^{\infty}v_{0,0}^{(i)}\epsilon^i.\label{eq1}
\end{equation}
We just look at the first equation of the BT. This reads
\[
\sum_{i=1}^{\infty}v_{0,0}^{(i)}\epsilon^i\left(\alpha u_{2,0}+\alpha\sum_{i=1}^{\infty}v_{1,0}^{(i)}\epsilon^{i}-\alpha u_{0,0}-\epsilon u_{0,0}\right)=\epsilon u_{1,0}\left(u_{0,0}+u_{2,0}+\sum_{i=1}^{\infty}v_{1,0}^{(i)}\epsilon^i\right).
\]
The leading order approximation gives
\[v_{0,0}^{(1)}=\frac{u_{1,0}(u_{0,0}+u_{2,0})}{\alpha(u_{2,0}-u_{0,0})}.\]
Higher order terms give
\begin{equation}
v_{0,0}^{(i)}=\frac{u_{0,0}v_{0,0}^{(i-1)}+u_{1,0}v_{1,0}^{(i-1)}-\alpha\sum_{j=1}^{i-1}v_{0,0}^{(i)}v_{0,0}^{(j-i)}}{\alpha(u_{2,0}-u_{0,0})},~~~i=2,3....
\end{equation}
As in the case of the dKdV equation all these formulas are on the horizontal line, i.e. all the $v_{0,0}^{(i)}$ only depend
on values of $u_{i,j}$ with $j=0$.
An infinite sequence of conservation laws can be
obtained starting from the $\epsilon$ expansion of a single conservation law. It is straightforward to check that if we define
\begin{equation}
F=\ln\left(\frac{u_{0,0,1}}{u_{0,0,0}}\right),~~~G=-\ln\left(\frac{\alpha u_{0,0,1}-\theta u_{1,0,0}}{u_{0,0,0}}\right),\label{eq2}
\end{equation}
then 
\[(S_k-I)F+(S_l-I)G=0,\]
on solutions of (\ref{sys10}).
By plugging (\ref{eq1}) and $\theta=\alpha + \epsilon$ into (\ref{eq2}) and expanding $F=\sum_{i=0}^{\infty}F_i\epsilon^i$ and $G=-\ln(2\epsilon)+\sum_{i=0}^{\infty}G_i\epsilon^i$  we obtain 
\begin{align*}
&F_0=\ln\left(\frac{u_{1,0}}{u_{0,0}}\right),&~~~&F_1=\frac{u_{2,0}+u_{0,0}}{u_{2,0}-u_{0,0}},\\
&G_0=-\ln\left(\frac{u_{1,0}}{u_{2,0}-u_{0,0}}\right),&~~~&G_1=\frac{u_{1,0}(3u_{2,0}+u_{0,0})+u_{3,0}(u_{2,0}-u_{0,0})}{2u_{0,0}(u_{3,0}-u_{0,0})(u_{2,0}-u_{0,0})},\end{align*}
\begin{align*}
&F_2=\frac{4u_{2,0}u_{0,0}(u_{3,0}+u_{1,0})+(u_{2,0}^2-u_{0,0}^2)(u_{3,0}-u_{1,0})}{2(u_{2,0}-u_{0,0})^2(u_{1,0}-u_{3,0})},\\
&G_2=\frac{(u_{3,0}u_{2,0}+3u_{2,0}u_{1,0}+u_{1,0}u_{0,0}-u_{3,0}u_{0,0})^2}{8(u_{2,0}-u_{0,0})^2(u_{3,0}-u_{1,0})^2},~etc.
\end{align*}
Thus we see how the expansion of the BT around the point $\theta=\alpha$ yields an infinite sequence of conservation laws on the horizontal line.
Expansion around $\theta=\beta$ does not seem to yield an infinite sequence of conservation laws on the vertical
line, since (\ref{as0}) is asymmetric. In the case of the dKdV equation expansion around $\theta=\beta$ also gives an infinite number of conservation laws. 
\section{The Gardner method for conservation laws of integrable equations on
the quad-graph}
In this section we present all the necessary information for the application of the Gardner method to the quad-graphs that are listed in \cite{ABS}. The general form of ABS equations is
\begin{equation}
P_1~:~P(u_{0,0,0},u_{1,0,0},u_{0,1,0},u_{1,1,0},\alpha,\beta)=0.
\label{eqg1}
\end{equation}
For the application of the Gardner method we need a special BT and the initial conservation law. 
For all the ABS equations the necessary BTs are known and these are so called natural auto-BTs \cite{At0}. The general form of the natural auto-BTs for ABS equations can be obtained from the form of the equation (\ref{eqg1}). For (\ref{eqg1}) the form of the natural auto-BT is
\begin{eqnarray}
P_2&:&P(u_{0,0,0},u_{1,0,0},u_{0,0,1},u_{1,0,1},\alpha,\theta)=0,\label{eqb1}\\
P_3&:&P(u_{0,0,0},u_{0,0,1},u_{0,1,0},u_{0,1,1},\theta,\beta)=0.\label{eqb2}
\end{eqnarray}
This is the BT which we need for the application of the Gardner method.
Note that equations (\ref{eqg1},\ref{eqb1},\ref{eqb2}) are one equation embedded in three dimensions. 

The initial conservation law(ICL) also follows from multidimensional consistency. 
Expressions for conservation laws for $P_1,~P_2,~P_3$ are respectively 
\begin{eqnarray}
(S_k-I)F_1+(S_l-I)G_1=0,\nonumber\\
(S_k-I)F_2+(S_m-I)H_2=0,\label{y40}\\
(S_l-I)G_3+(S_m-I)H_3=0.\nonumber
\end{eqnarray} 
Five-point conservation laws for ABS equations were found in \cite{RH3}, where the authors show that each ABS equation has three five-point conservation
laws. Two five-point conservation laws for $P_1$ can be always presented as
\begin{eqnarray}
F=f(u_{0,-1,0},u_{0,0,0},u_{0,1,0},\beta),&~~~&G=g(u_{0,-1,0},u_{0,0,0},u_{1,0,0},\alpha,\beta),\\
F'=g(u_{-1,0,0},u_{0,0,0},u_{0,1,0},\beta,\alpha),&~~~&G'=f(u_{-1,0,0},u_{0,0,0},u_{1,0,0},\alpha).
\end{eqnarray}
As we said before, equations $P_1,~P_2,~P_3$ are one equation which is embedded in three dimensions. Therefore conservation laws for $P_1,~P_2$ can be obtained from conservation laws for $P_1$ by rewriting them on the appropriate plane.
The conservation law for $P_2$ which corresponds to $F,~G$ is
\begin{equation}
F_2=f(u_{0,0,-1},u_{0,0,0},u_{0,0,1},\theta),~~~H_2=g(u_{0,0,-1},u_{0,0,0},u_{1,0,0},\alpha,\theta).\label{CL1}
\end{equation}
The conservation law for $P_3$ which corresponds to $F,~G$ is
\begin{equation}
G_3=f(u_{0,0,-1},u_{0,0,0},u_{0,0,1},\theta),~~~H_3=g(u_{0,0,-1},u_{0,0,0},u_{0,1,0},\beta,\theta).\label{CL2}
\end{equation}
$F_2$ is equal to $G_3$, so with the help of linear combination of expressions from (\ref{y40}) we obtain that
\begin{multline*}
(S_l-I)(S_k-I)F_2+(S_l-I)(S_m-I)H_2-(S_k-I)(S_l-I)G_3\\-(S_k-I)(S_m-I)H_3=(S_m-I)((S_l-I)H_2-(S_k-I)H_3)=0
\end{multline*}
is true on solutions of $P_2$ and $P_3$. This expression can be integrated with respect to $m$ to give
\begin{equation}
(S_l-I)H_2-(S_k-I)H_3=C(k,l)\label{ICL1}
\end{equation}
where $C(k,l)$ is a function which has to be determined. 
Equations $P_2$ and $P_3$ are symmetric with respect to the reflection of the coordinate $m$, that is
\begin{eqnarray*}P(u_{0,0,0},u_{1,0,0},u_{0,0,1},u_{1,0,1},\alpha,\theta)&=&\pm P(u_{0,0,1},u_{1,0,1},u_{0,0,0},u_{1,0,0},\alpha,\theta)\\&=&\pm S_mP(u_{0,0,0},u_{1,0,0},u_{0,0,-1},u_{1,0,-1},\alpha,\theta).\end{eqnarray*}
Therefore (\ref{ICL1}) is also symmetric with respect to the reflection of coordinate $m$, so
\begin{equation}
(S_l-I)g(u_{0,0,1},u_{0,0,0},u_{1,0,0},\alpha,\theta)-(S_k-I)g(u_{0,0,1},u_{0,0,0},u_{0,1,0},\beta,\theta)=C(k,l).\label{ICL2}
\end{equation}
By substitution of $P_2$ and $P_3$ into (\ref{ICL2}) we obtain that $C(k,l)=0$ for all the ABS equations.
Therefore 
\[
(S_l-I)H_2-(S_k-I)H_3=0,
\]
for all the ABS equations.
Thus, the components of ICLs for all the ABS equations are
\[
F_{ICL}=-g(u_{0,0,1},u_{0,0,0},u_{0,1,0},\alpha,\theta),~~~G_{ICL}=g(u_{0,0,1},u_{0,0,0},u_{1,0,0},\beta,\theta).
\]
We present a table with ICLs for all the ABS equations bellow.
The equations from the ABS
classification are as follows; for convenience, we have used the
form of $\mathbf{Q4}$ that was discovered by Hietarinta \cite{Hie0}.
{\linespread{1.3}
\begin{footnotesize}
\begin{eqnarray}
\mathbf{Q1}:~~&\alpha(u_{0,0}-u_{0,1})(u_{1,0}-u_{1,1})-\beta(u_{0,0}-u_{1,0})(u_{0,1}-u_{1,1})+\delta^2\alpha\beta(\alpha-\beta)=0,\nonumber\\
\mathbf{Q2}:~~&\alpha(u_{0,0}-u_{0,1})(u_{1,0}-u_{1,1})-\beta(u_{0,0}-u_{1,0})(u_{0,1}-u_{1,1})\nonumber\\&+\alpha\beta(\alpha-\beta)(u_{0,0}+u_{1,0}+u_{0,1}+u_{1,1} )-\alpha\beta ( \alpha-\beta ) ( {\alpha}^2-\alpha\beta+{\beta}^2 ) =0,\nonumber\\
\mathbf{Q3}:~~&({\beta}^2-{\alpha}^2)(u_{0,0}u_{1,1}+u_{1,0}u_{0,1})+\beta({\alpha}^2-1)(u_{0,0}u_{1,0}+u_{0,1}u_{1,1})\nonumber\\&-\alpha({\beta}^2-1)(u_{0,0}u_{0,1}+u_{1,0}u_{1,1})-\delta^2{{({\alpha}^2-{\beta}^2)({\alpha}^2-1)({\beta}^2-1)}/(4\alpha\beta})=0,\nonumber\\
\mathbf{Q4}:~~&\mathrm{sn}(\alpha)(u_{0,0}u_{1,0}+u_{0,1}u_{1,1})-\mathrm{sn}(\beta)(u_{0,0}u_{0,1}+u_{1,0}u_{1,1})-\mathrm{sn}(\alpha-\beta)(u_{0,0}u_{1,1}+u_{1,0}u_{0,1})\nonumber\\&+\mathrm{sn}(\alpha-\beta)\mathrm{sn}(\alpha)\mathrm{sn}(\beta)(1+K^2u_{0,0}u_{1,0}u_{0,1}u_{1,1})=0,\label{ABS}\\
\mathbf{H1}:~~&(u_{0,0}-u_{1,1} ) ( u_{1,0}-u_{0,1} ) +\beta-\alpha=0,\nonumber\\
\mathbf{H2}:~~&(u_{0,0} -u_{1,1})(u_{1,0}-u_{0,1})+(\beta-\alpha)(u_{0,0}+u_{1,0}+u_{0,1}+u_{1,1})+{\beta}^2-{\alpha}^2=0,\nonumber\\
\mathbf{H3}:~~&\alpha(u_{0,0}u_{1,0}+u_{0,1}u_{1,1})-\beta(u_{0,0}u_{0,1}+u_{1,0}u_{1,1})+\delta^2({\alpha}^2-{\beta}^2)=0,\nonumber\\
\mathbf{A1}:~~&\alpha(u_{0,0}+u_{0,1})(u_{1,0}+u_{1,1})-\beta(u_{0,0}+u_{1,0})(u_{0,1}+u_{1,1})-\delta^2\alpha\beta(\alpha-\beta)=0,\nonumber\\
\mathbf{A2}:~~&({\beta}^2-{\alpha}^2)(u_{0,0}u_{1,0}u_{0,1}u_{1,1}+1)+\beta({\alpha}^2-1)(u_{0,0}u_{0,1}+u_{1,0}u_{1,1})\nonumber\\
&-\alpha({\beta}^2-1)(u_{0,0}u_{1,0}+u_{0,1}u_{1,1})=0.\nonumber
\end{eqnarray}\end{footnotesize}}
Here $\mathrm{sn}(\alpha)=\mathrm{sn}(\alpha;K)$ is a
Jacobi elliptic function with modulus $K$. Without loss of
generality, the parameter $\delta$ is restricted to the values $0$
and $1$.
Obtained ICLs are summarized in table 1, in which we list the components $F$ and $G$ for each of the ABS equations.

\begin{longtable}{ll}
\caption{\bfseries Initial conservation laws for the equations from
the ABS classification}\\

{\bfseries Eq.}& {\bfseries Components} \\ \hline
\endfirsthead
{\bfseries Eq.}& {\bfseries Components} \\ \hline
\endhead
\hline \multicolumn{2}{r}{\emph{Continued on next page}}
\endfoot
\hline
\endlastfoot
\\

$\mathbf{Q1}$
&$\begin{array}{l}
F=\ln\left({-\tilde{u}_{0,0}+ u_{0,0}-\delta\theta})-\ln({\delta(\beta-\theta)+u_{0,1}-\tilde{u}_{0,0}}\right),\\
G=-\ln\left({u_{0,0}-\tilde{u}_{0,0}-\delta\theta})+\ln({\delta(\alpha-\theta)+u_{1,0}-\tilde{u}_{0,0}}\right),
\end{array}$
\\&\\
$\mathbf{Q2}$
&$\begin{array}{l}
F=\ln({(\theta\beta^2+\beta(\tilde{u}_{0,0}-u_{0,0}-\theta^2)+\theta(u_{0,0}-u_{0,1}))^2})\\
-\ln({\beta^4-2\beta^2(u_{0,1}+u_{0,0})+(u_{0,0}-u_{0,1})^2}),\\
G=-\ln({(\theta\alpha^2+\alpha(\tilde{u}_{0,0}-u_{0,0}-\theta^2)+\theta(u_{0,0}-u_{1,0}))^2})\\
+\ln({\alpha^4-2\alpha^2(u_{1,0}+u_{0,0})+(u_{0,0}-u_{1,0})^2}),
  \end{array}$\\&\\
$\mathbf{Q3}$
&$\begin{array}{l}
F=\ln({(\theta^2(u_{0,0}-\beta u_{0,1}))^2-\theta(1-\alpha^2)\tilde{u}_{0,0}-\alpha^2u_{0,0}+\alpha^2u_{0,1}})\\-\ln({4\beta(\beta u_{0,0}-u_{0,1})(u_{0,0}-\beta u_{0,1})+\delta(1-\beta^2)^2}),\\
G=-\ln({(\theta^2(u_{0,0}-\alpha u_{1,0})-\theta\tilde{u}_{0,0}(1-\alpha^2)-\alpha^2u_{0,0}+\alpha u_{1,0})^2})\\+\ln({4\alpha(\alpha u_{0,0}-u_{1,0})(u_{0,0}-\alpha u_{1,0})+\delta(1-\alpha^2)^2}),
  \end{array}$\\&\\
$\mathbf{Q4}$&$
\begin{array}{l}
F=\ln({\mathrm{sn}(\theta)^2(1+K^2u_{0,0}\tilde{u}_{0,0})+2\mathrm{cn}(\theta)\mathrm{dn}(\theta)u_{0,0}\tilde{u}_{0,0}-u_{0,0}^2-\tilde{u}_{0,0}^2})\\-\ln(\mathrm{sn}(\theta-\beta)^2(u_{0,1}\tilde{u}_{0,0}-\mathrm{sn}(\theta)\mathrm{sn}(\beta))(\mathrm{sn}(\theta)\mathrm{sn}(\beta)K^2\tilde{u}_{0,0}u_{0,1}-1)\\-(\mathrm{sn}(\theta)u_{0,1}-\mathrm{sn}(\beta)\tilde{u}_{0,0})(\mathrm{sn}(\beta)u_{0,1}-\mathrm{sn}(\theta)\tilde{u}_{0,0})),\\
G=-\ln({\mathrm{sn}(\theta)^2(1+K^2u_{0,0}\tilde{u}_{0,0})+2\mathrm{cn}(\theta)\mathrm{dn}(\theta)u_{0,0}\tilde{u}_{0,0}-u_{0,0}^2-\tilde{u}_{0,0}^2})\\+\ln(\mathrm{sn}(\alpha-\theta)^2(u_{1,0}\tilde{u}_{0,0}-\mathrm{sn}(\theta)\mathrm{sn}(\alpha))(\mathrm{sn}(\theta)\mathrm{sn}(\alpha)K^2\tilde{u}_{0,0}u_{1,0}-1)\\-(\mathrm{sn}(\theta)u_{1,0}-\mathrm{sn}(\alpha)\tilde{u}_{0,0})(\mathrm{sn}(\alpha)u_{1,0}-\mathrm{sn}(\theta)\tilde{u}_{0,0})),
\end{array}$\\&\\
$\mathbf{H1}$
&$\begin{array}{l}
F=-\ln(\tilde{u}_{0,0}-u_{0,1}),\\
G=\ln(\tilde{u}_{0,0}-u_{1,0}),
\end{array}$
\\&$\begin{array}{l}
\end{array}
$\\&\\
$\mathbf{H2}$
&$\begin{array}{l}
F=\ln\left({(\beta-\theta+u_{0,1}-\tilde{u}_{0,0})^2})-\ln({\beta+u_{0,0}+u_{0,1}}\right),\\
G=-\ln\left({(\alpha-\theta+u_{1,0}-\tilde{u}_{0,0})^2})+\ln({\alpha+u_{0,0}+u_{1,0}}\right),
\end{array}$\\&\\
$\mathbf{H3}$&$\begin{array}{l}
F=\ln\left({(\beta \tilde{u}_{0,0}-\theta u_{0,1})^2})-\ln({\delta \beta+u_{0,0}u_{0,1}}\right),\\
G=-\ln\left({(\alpha \tilde{u}_{0,0}-\theta u_{1,0})^2})+\ln({\delta\alpha+u_{0,0}u_{1,0}}\right),
\end{array}$\\&\\
$\mathbf{A1}$
&$\begin{array}{l}
F=\ln((u_{0,0}+\tilde{u}_{0,0})^2-\delta\theta^2)-\ln((u_{0,1}-\tilde{u}_{0,0})^2-\delta(\theta-\beta)^2),\\
G=-\ln((u_{0,0}+\tilde{u}_{0,0})^2-\delta\theta^2)+\ln((u_{1,0}-\tilde{u}_{0,0})^2-\delta(\alpha-\theta)^2),
\end{array}$\\&\\
$\mathbf{A2}$
&$\begin{array}{l}
F=\ln(\tilde{u}_{0,0}u_{0,0}-\theta)(\theta\tilde{u}_{0,0}u_{0,0}-1)-\ln((\theta u_{0,1}-\beta \tilde{u}_{0,0})(\beta u_{0,1}-\theta \tilde{u}_{0,0})),\\
G=-\ln(\tilde{u}_{0,0}u_{0,0}-\theta)(\theta\tilde{u}_{0,0}u_{0,0}-1)+\ln((\theta u_{1,0}-\alpha \tilde{u}_{0,0})(\alpha u_{1,0}-\theta \tilde{u}_{0,0})),
\end{array}$\label{t1}\end{longtable}

Another problem which can appear during the application of the Gardner method is that $v_{0,0}^{(i)}$ has to be found for $i=1,2,...$ and for all the ABS equations. We prove that it is always possible to explicitly find $v_{0,0}^{(i)}$ for $i=1,2,...$ for all the ABS equations in the application of the Gardner method in the $k$ direction. For the $l$ direction the proof is similar.
Consider $\theta=\alpha+\epsilon$ where $\epsilon$ is small, and look for a solution of BT in the form
\begin{equation}
u_{0,0,1}=u_{1,0}+\sum_{i=1}^{\infty}v_{0,0}^{(i)}\epsilon^i.\label{kgb1}
\end{equation}
We look only at the first equation of BT. This reads
\begin{equation}
P\left(u_{0,0},u_{1,0},u_{1,0}+\sum_{i=1}^{\infty}v_{0,0}^{(i)}\epsilon^i,u_{2,0}+\sum_{i=1}^{\infty}v_{1,0}^{(i)}\epsilon^{i},\alpha,\alpha+\epsilon\right)=0.\label{y10}
\end{equation}
Let $C_n$ be the coefficient next to the $\epsilon^n$ in the expansion of (\ref{y10}) in the Taylor series around $\epsilon=0$. At first sight $C_n$ depends upon $v_{0,0}^{(i)}, v_{1,0}^{(i)},~i=1,2,...n$ and to find $v_{0,0}^{(i)}$ one has to solve non-trivial difference equations. We show that that $C_n$ depends just upon $v_{0,0}^{(i)}, v_{1,0}^{(j)},~i=1,2,...n,~j=1,2,...n-1$, so $C_n$ can be solved explicitly with respect to $v_{0,0}^{(n)}$. First of all let us show
\[
C_0=P\left(u_{0,0},u_{1,0},u_{1,0},u_{2,0},\alpha,\alpha\right)=0
\]
Since $P$ is an ABS equation it has the symmetry property \cite{ABS}
\[
P(x, u, v, y, \alpha, \beta)=-P(x, v,u, y,\alpha, \beta).
\]
According to this symmetry property we obtain
\[
P\left(u_{0,0},u_{1,0},u_{1,0},u_{2,0},\alpha,\alpha\right)=-P\left(u_{0,0},u_{1,0},u_{1,0},u_{2,0},\alpha,\alpha\right)=0.
\]
The derivative of (\ref{y10}) with respect $\epsilon$ is
\begin{equation}
\frac{d}{d\epsilon}P=\sum_{i=1}^{\infty}iv_{0,0}^{(i)}\epsilon^{i-1}P_3+\sum_{i=1}^{\infty}iv_{1,0}^{(i)}\epsilon^{i-1}P_4+P_6,\label{y30}
\end{equation}
where $P_i$ is the derivative of $P$ with respect to its $i$th argument. For example
\begin{equation*}
P_3=\left.\frac{\partial}{\partial a}P\left(u_{0,0},u_{1,0},a,u_{2,0}+\sum_{i=1}^{\infty}v_{1,0}^{(i)}\epsilon^{i},\alpha,\alpha+\epsilon\right)\right|_{a=u_{1,0}+\sum_{i=1}^{\infty}v_{0,0}^{(i)}\epsilon^i}.
\end{equation*}
For $\epsilon=0$ we obtain $C_1$
\begin{multline}
C_1=\left.\frac{d}{d\epsilon}P\right|_\epsilon=v_{0,0}^{(1)}P_3\left(u_{0,0},u_{1,0},u_{1,0},u_{2,0},\alpha,\alpha\right)+v_{1,0}^{(1)}P_4\left(u_{0,0},u_{1,0},u_{1,0},u_{2,0},\alpha,\alpha\right)\\+P_6\left(u_{0,0},u_{1,0},u_{1,0},u_{2,0},\alpha,\alpha\right).\label{y20}
\end{multline}
We checked that for all the ABS equations
\[P_3\left(u_{0,0},u_{1,0},u_{1,0},u_{2,0},\alpha,\alpha\right)\neq0,~~~P_4\left(u_{0,0},u_{1,0},u_{1,0},u_{2,0},\alpha,\alpha\right)=0.\]
So, $C_1$ depends upon $v_{0,0}^{(1)}$ and does not depend upon $v_{1,0}^{(1)}$. From (\ref{y30}) it is seen that generally the form of $C_n$ is similar to the form of $C_1$, namely
\[
C_n=n!v_{0,0}^{(n)}P_3\left(u_{0,0},u_{1,0},u_{1,0},u_{2,0},\alpha,\alpha\right)+n!v_{1,0}^{(n)}P_4\left(u_{0,0},u_{1,0},u_{1,0},u_{2,0},\alpha,\alpha\right)+R
\]
Here $R$ does not depend upon $v_{0,0}^{(n)},~v_{1,0}^{(n)}$. As we said before $P_4\left(u_{0,0},u_{1,0},u_{1,0},u_{2,0},\alpha,\alpha\right)=0$ therefore $C_n$ can be solved with respect to $v_{0,0}^{(n)}$. By solving $C_i$ with respect to $v_{0,0}^{(i)}$ iteratively for $i=1,2,...$ we can find as many $v_{0,0}^{(i)}$ as necessary. Once $v_{0,0}^{(i)},~i=1...n$ are known, we plug (\ref{kgb1}) into the initial conservation law and expand it around $\epsilon=0$ up to order $n$. In this way $n$ conservation laws can be found.
\section{The Gardner method for the symmetries of integrable equations on
the quad-graph}
In \cite{RH2} five-point symmetries were found for ABS equations. 
The authors show there that each ABS equation $P_1$ has four five-point symmetries. Two of these symmetries can always be presented as
\begin{eqnarray}
X_h&=&\eta(u_{-1,0,0},u_{0,0,0},u_{1,0,0},\alpha)\partial_{u_{0,0,0}},\\
X_v&=&\eta(u_{0,-1,0},u_{0,0,0},u_{0,1,0},\beta)\partial_{u_{0,0,0}}.
\end{eqnarray}
Therefore equations $P_2$ and $P_3$ both have a symmetry
\begin{equation}
X=\eta(u_{0,0,-1},u_{0,0,0},u_{0,0,1},\theta)\partial_{u_{0,0,0}}.
\end{equation}
Since $P_1,~P_2$ and $P_3$ are consistent in three dimensions we obtain that $X$ is also a symmetry for $P_1$. Let us call $X$ the \textit{initial symmetry} for $P_1$.
Let us apply the Gardner method to $X$ or, in other words, expand $X$ in a series. $X$ involves the BT and inverse BT $u_{0,0,1}$ and $u_{0,0,-1}$, therefore we are looking for the solution of those for $\theta=\alpha\pm\epsilon$ where $\epsilon$ is small in the forms 
\begin{eqnarray}
u_{0,0,1}&=&u_{1,0,0}+\sum_{i=1}^{\infty}v_{0,0}^{(i)}\epsilon^i,\\
u_{0,0,-1}&=&u_{-1,0,0}+\sum_{i=1}^{\infty}w_{0,0}^{(i)}\epsilon^i.
\end{eqnarray}
By plugging this in the characteristic $\eta$ of the symmetry $X$ we obtain
\begin{equation}
\eta=\eta(u_{-1,0,0}+\sum_{i=1}^{\infty}w_{0,0}^{(i)}\epsilon^i,u_{0,0,0},u_{1,0,0}+\sum_{i=1}^{\infty}v_{0,0}^{(i)}\epsilon^i,\alpha+\epsilon).\label{sym1}
\end{equation}
After expansion of $\eta$ in Taylor series around $\epsilon=0$ we obtain
\begin{equation}
\eta=\sum_{i=1}^{\infty}\eta_i\epsilon^i.\label{sym2}
\end{equation}
Each $\eta_i,~i=1,2,...$ is the characteristic of a symmetry for $P_1$.
For example let us consider the discrete KdV equation
\begin{equation}(u_{0,0,0}-u_{1,1,0} ) ( u_{1,0,0}-u_{0,1,0} ) +\beta-\alpha=0.\label{eq01}
\end{equation}
Two of the symmetries for this equation are
\begin{eqnarray}
X_h&=&\frac{1}{u_{1,0,0}-u_{-1,0,0}}\partial_{u_{0,0,0}},\\
X_v&=&\frac{1}{u_{0,1,0}-u_{0,-1,0}}\partial_{u_{0,0,0}}.
\end{eqnarray}
Therefore
\begin{equation}
X=\frac{1}{u_{0,0,1}-u_{0,0,-1}}\partial_{u_{0,0,0}}.
\end{equation}
is also a symmetry.
Expression (\ref{sym1}) for $X$ is 
\[\eta=\frac{1}{u_{1,0,0}-u_{-1,0,0}+\sum_{i=1}^{\infty}(v_{0,0}^{(i)}-w_{0,0}^{(i)})\epsilon^i}. \]
We just look at the first equation of BT which is $P_2$. This reads
\begin{eqnarray}
\epsilon&=&\left(\sum_{i=1}^{\infty}v_{0,0}^{(i)}\epsilon^i\right)\left(u_{0,0,0}-u_{2,0,0}-\sum_{i=1}^{\infty}v_{1,0}^{(i)}\epsilon^i\right),\\
\epsilon&=&\left(\sum_{i=1}^{\infty}w_{1,0}^{(i)}\epsilon^i\right)\left(u_{-1,0,0}-u_{1,0,0}+\sum_{i=1}^{\infty}w_{0,0}^{(i)}\epsilon^i\right).
\end{eqnarray}
The leading order approximation gives 
\begin{eqnarray} 
v^{(1)}_{0,0} = \frac1{u_{0,0,0}-u_{2,0,0}},\\  
w^{(1)}_{1,0} = \frac1{u_{-1,0,0}-u_{1,0,0}}.
\end{eqnarray}
Higher order terms give 
\begin{eqnarray}   
v^{(i)}_{0,0}&=& \frac1{u_{0,0,0}-u_{2,0,0}} \sum_{j=1}^{i-1} v^{(j)}_{0,0} v^{(i-j)}_{1,0}\ , \qquad i=2,3,\ldots \ ,\\
w^{(i)}_{1,0}&=& -\frac1{u_{-1,0,0}-u_{1,0,0}} \sum_{j=1}^{i-1} w^{(j)}_{0,0} w^{(i-j)}_{1,0}\ , \qquad i=2,3,\ldots .
\end{eqnarray}
The first three coefficients in the expansion of $\eta$ around $\epsilon=0$ are
\begin{eqnarray*} 
\eta_1&=&\frac{1}{u_{1,0}-u_{-1,0}},\\  
\eta_2&=&\frac{u_{-2,0}-u_{2,0}}{(u_{-1,0}-u_{1,0})^2(u_{0,0}-u_{2,0})(u_{0,0}-u_{-2,0})},\\  
\eta_3&=&\frac{u_{-1,0}-u_{3,0}}{(u_{0,0}-u_{2,0})^2(u_{1,0}-u_{3,0})(u_{-1,0}-u_{1,0})^3}+\frac{2}{(u_{0,0}-u_{2,0})(u_{-2,0}-u_{2,0})(u_{-1,0}-u_{1,0})^3}\\
&+&\frac{u_{-3,0}-u_{1,0}}{(u_{0,0}-u_{-2,0})^2(u_{-3,0}-u_{-1,0})(u_{-1,0}-u_{1,0})^3}-\frac{2}{(u_{0,0}-u_{-2,0})(u_{-2,0}-u_{2,0})(u_{-1,0}-u_{1,0})^3}.
\end{eqnarray*}
These are characteristics of symmetries for (\ref{eq01}). $\eta_1$ and $\eta_2$ were already presented in \cite{TTP}.
\section{Concluding remarks}
In this article we showed how to apply the Gardner method for generation of conservation laws to all the ABS equation. It is shown that all the necessary information for the application of the Gardner method follows from the multidimensional consistency of ABS equations. Namely the natural auto-BT for ABS equation is the equation itself embedded in three dimensions. This construction is possible because of the multidimensional consistency of the equation. The initial conservation law for ABS equations follows from the linear combinations of conservation laws for the corresponding embedded equations.

Another important result of this article is the introduction of the Gardner method for symmetry generation for ABS equations. 
In the Gardner method for the symmetry generation one has to substitute expressions
\[
u_{0,0,1}=u_{1,0,0}+\sum_{i=1}^{\infty}v_{0,0}^{(i)}\epsilon^i,~~~u_{0,0,-1}=u_{-1,0,0}+\sum_{i=1}^{\infty}w_{0,0}^{(i)}\epsilon^i.
\]
into the initial symmetry. The expansion of this symmetry in a Taylor series gives an infinite number of symmetries.

Application of the Gardner method for generation of conservation laws for the asymmetric equation
\begin{equation}
\alpha u_{0,0}u_{1,0}-\beta(u_{0,0}u_{0,1}+u_{1,0}u_{1,1})=0.\label{kgb2}
\end{equation}
is also considered. In this case we can again say that all necessary information for application of the method follows from  multidimensional consistency.  Equation (\ref{kgb2}) is consistent with the symmetric equation
\begin{equation}
\alpha( u_{0,0}u_{1,0}+u_{0,1}u_{1,1})-\beta(u_{0,0}u_{0,1}+u_{1,0}u_{1,1})=0.\label{kgb3}
\end{equation}
We obtain a three dimensional consistent system by imposing (\ref{kgb3}) on 2 opposite sides of a cube and (\ref{kgb2}) on the remaining 4 sides.
From this consistency follows the Lax pair for (\ref{kgb2})
\[L=\left[ \begin
{array}{cc}1&\alpha\,u_{1,0}\\\noalign{\medskip}{\displaystyle\frac
{\alpha{\lambda}^{2}}{u_{0,0}}}&
\displaystyle-\frac{u_{1,0}}{u_{0,0}}\end {array} \right],~~
M=\left[ \begin {array}{cc}
1&\beta\,u_{0,1}\\\noalign{\medskip}{\displaystyle\frac
{\beta\lambda^{2}}{u_{0,0}}}&0\end {array} \right].\]
Equation (\ref{kgb2}) satisfies all the criteria of integrability \cite{Za0}, namely there exist infinite number of conservation laws and symmetries, and a Lax pair. As
is seen from this article all these properties are connected with multidimensional consistency. So, multidimensional consistency is a universal criterion of integrability. It is not necessary for an equation to be consistent with itself, as we showed for (\ref{kgb2}) it can be consistent with other equations. 

Here are some interesting topics for future research:
\begin{itemize}
\item The conditions for the application of the Gardner method to the consistent equation (equation and its BT).
\item The classification of quad-graph equations according to these conditions.  
\end{itemize}
\bibliographystyle{acm} \bibliography{P}
\end{document}